# Characterization and Identification of Au Pathfinder Minerals from an Artisanal Mine Site using X-Ray Diffraction

Gabriel Nzulu, Per Eklund and Martin Magnuson

*Department of Physics, Chemistry and Biology (IFM) Linköping University, Sweden*

## Abstract

Gold associated pathfinder minerals have been investigated by identifying host minerals of Au for samples collected from an artisanal mining site near a potential gold mine (Kubi Gold Project) in Dunwka-On-Offin in the Central Region of Ghana. We find that for each composition of Au powder (impure) and the residual black hematite/magnetite sand that remains after gold panning, there is a unique set of associated diverse indicator minerals. These indicator minerals are identified as $SiO_2$ (quartz), $Fe_3O_4$ (magnetite) and $Fe_2O_3$ (hematite) while contributions from pyrite, arsenopyrites, iridosmine, scheelite, tetradymite, garnet, gypsum and other sulfate materials are insignificant. This constitutes a confirmative identification of Au pathfinding minerals in this particular mineralogical area. The findings suggest that x-ray diffraction could also be applied in other mineralogical sites to aid in identifying indicator minerals of Au and location of ore bodies at reduced environmental and exploration costs.

## 1. Introduction

At mineralogical mining sites, fast location of ore bodies is paramount in order to reduce exploration cost. For this purpose, pathfinding minerals are important. These minerals act as aid in the original ore-body discovery. Au that can be traced from the presence of pathfinding minerals mostly originate as anhedral crystal assemblies (*i.e.*, without well-defined crystal facets) that naturally exist as single or polycrystalline mineral aggregates that are usually found *in-situ* in hydrothermal quartz veins and other kinds of key deposits in metamorphic and igneous rocks [1, 2].

The most common mineral at most Au mining sites is pyrite ($FeS_2$) that can also be found in oil shales and coal [3]. Other common minerals at Au mining sites are arsenopyrite, different forms of silicates minerals (garnet), and magnetite ($Fe_3O_4$). Both mineralogical and geochemical information are indispensable to provide an initial valuation of the potential ore zone of an exploration area.

Quantitative interpretation of x-ray diffraction (XRD) data [4] have long been applied to distinguish between mineral assemblages, and to define chemical and mineralogical compositions [5]. Previous XRD studies of Au and associated minerals have mostly been performed to determine the grain size morphology and crystallinity [6]. XRD has been used to conclude that highly hydrated and water saturated environments contribute to the migration of Au within alluvia regimes and on hydrothermal mineral assemblages [7-9]. Multivariate statistical analysis and geostatistical methods have been applied to identify pathfinding elements [10-11]. Bayari *et al.* [4] found that mineralized regolith profiles and mobility of elements (minerals) in the soil at the Bole-Nangoli gold belt in the north-eastern Ghana could mainly be attributed to amorphous mineral phases. Furthermore, Zhao and Pring (2019) [12], studied the mineral transformation in Au and silver (Ag) in fluids using the telluride group of minerals associated with Au and focused on the texture, reaction mechanism and the kinetics of the oxidation leaching of the tellurides.

Cairns *et al.* [13] identified topsoil minerals and pathfinders of Au by considering the fine grain size and amorphous nature of the minerals. Furthermore, XRD studies on the influence of thermal stability of magnetite investigate the effect of temperature on the phase transitions [14-21]. This information is of importance for the investigation of magnetite as a pathfinder mineral of Au. As follows from this background, there is still a need for characterization of pathfinding Au-associated mineral by XRD on residual samples to establish their relationship and to preserve information about the physiochemical situations of their origin.

In this work, we investigate the crystal structure of Au in relation to the corresponding pathfinding minerals such as quartz ($SiO_2$), $Fe_3O_4$, $Fe_2O_3$, $FeS_2$ and $Fe_{1-x}S$, collected from an artisanal mining site, *i.e.*, a small scale hand-mining site, in the central region of Ghana. XRD was used for phase identification and to obtain structural information including Rietveld refinement. In addition to the known minerals, we also identified hematite ($Fe_2O_3$) as an important pathfinding mineral. The present study can be used to enable future identification of pathfinding minerals for Au exploration.

## 2. Experimental details

### 2.1 Description of the Field Site

The sample collection site is located close to the Kubi Gold (Adansi Gold) on the outskirt of Dunkwa-On-Offin, (5° 58'11.32" N, 1°46' 59.15" W) as shown in Fig. 1. Dunkwa is the capital of the Upper Denkyira East Municipal District located in the central region of Ghana and is drained by several rivers and streams with the Offin River serving as the main river source. The location follows the geology of Ghana which is associated with antiquity of crystalline basement rock, volcanic belts, and sedimentary basins. Most Au is found in steeply dipping quartz veins in shear zones within the Birimian basins with sulfur rich minerals like arsenopyrites and $FeS_2$. Other sources of Au found are alluvia placer Au in the Offin river deposits in gravels as well as some mineralized placer Au reconstituted with minerals like $Fe_3O_4$ and $Fe_2O_3$ in quartz-pebble conglomerates of the Tarkwaian deposits [22]. Extremely oxidized, weathered or putrefied rock commonly located at the upper and exposed part of the ore deposit or mineralized vein known as "gossan" or iron cap serves as a guide to trace buried Au ore deposits in this area [23]. The surface oxides of the minerals at this site are usually red, orange to yellowish brown color serving as alteration to the parent rock or soil.

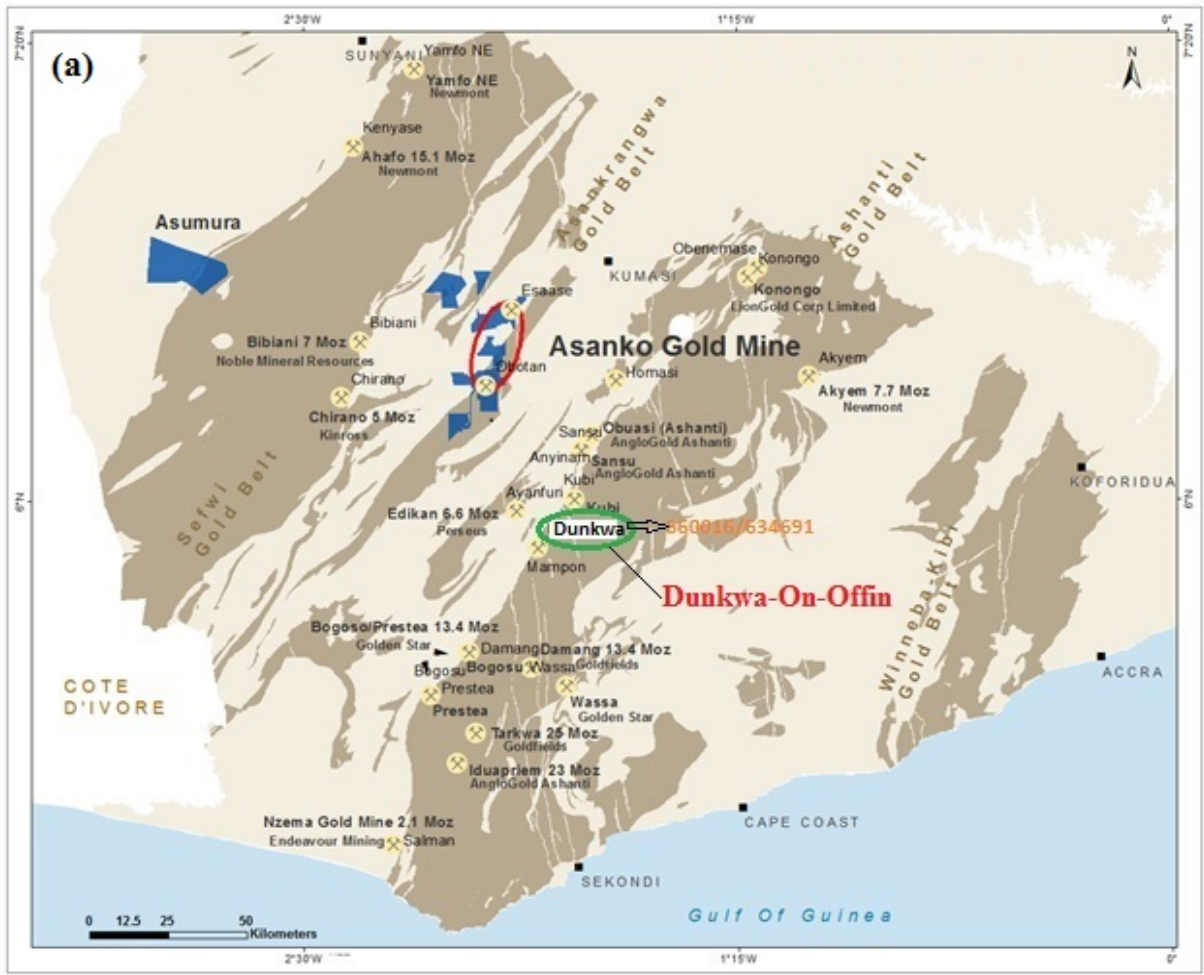

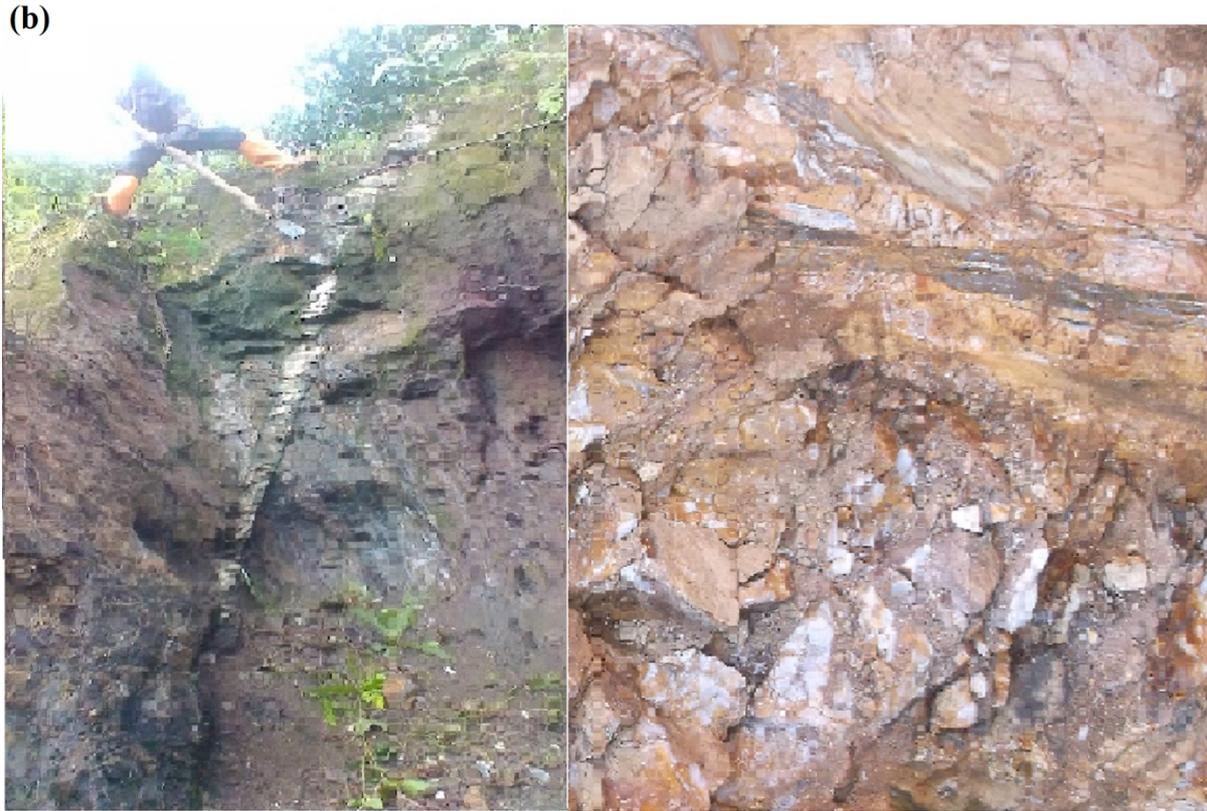

**Figure 1:** (a) Geological Map of Mining area in and around Dunkwa-On-Offin, in the Ashanti Gold Belt of the central Ghana [24] (b) Photograph of sampling collection area of the Artisanal Mining Site. On site photos by G. Nzulu in Nov. 2019.

## 2.2 Sample Preparation

Sediment samples that contain Au were extracted from a depth of 10 m of an artisanal mining site in Dunkwa-On-Offin. Figure 2 shows the depth profile. Each sample was divided into two parts where one portion was refined into pure solid Au whilst the other part of the powder sample was subjected to Au panning *i.e.*, washing and magnetic extraction of Fe-based minerals as shown in Fig. 2. The final three samples (Fig. 3) containing a solid Au nugget, untreated (impure) Au powder, and the separated black sand-like minerals were examined by X-ray diffraction. The size of the two powder crystal samples ranges from 0.05 cm to about 0.2 cm in maximum dimension of which most were hoppered single crystals with an octahedral crystal structure with a few being of non-octahedral forms.

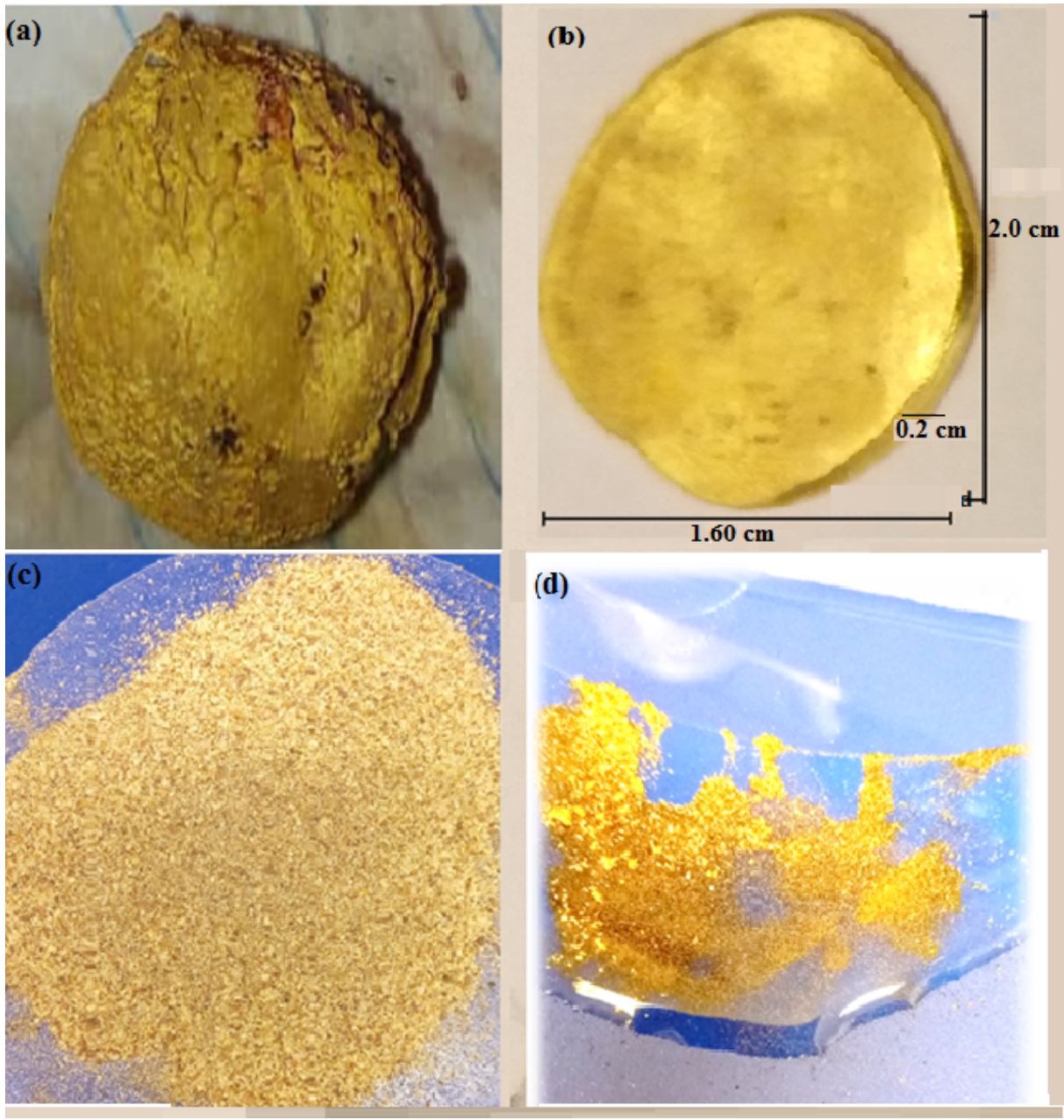

**Figure 2**: (a) Wet residual sample from Dunkwa-On-Offin artisanal mine site (b) Refined part of sample into a Au nugget 22 carats as measured with a digital electronic Au purity Analyzer DH 300K from VTSYIQI. (c) Dried sample after coarse rinsing before fine panning. Note that the sample contains white quartz as well as black magnetite and hematite. (d) Final impure Au after fine panning. Photos by G. Nzulu.

(hematite) sink to the bottom of the pan. While the black sand remained in the pan, a strong permanent magnet was swept over (to and fro) in a circular motion, a couple of centimeters above the material to maximize the magnetic susceptibility (induced ferromagnetics in the $Fe_2O_3$) for easy capture of the magnetite and hematite. The process was repeated until there was no more added material on the surface of the permanent magnet. The magnetically captured material was dominated by magnetite that is a pathfinding mineral (Fig. 3b) in addition minority minerals that can be identified using XRD.

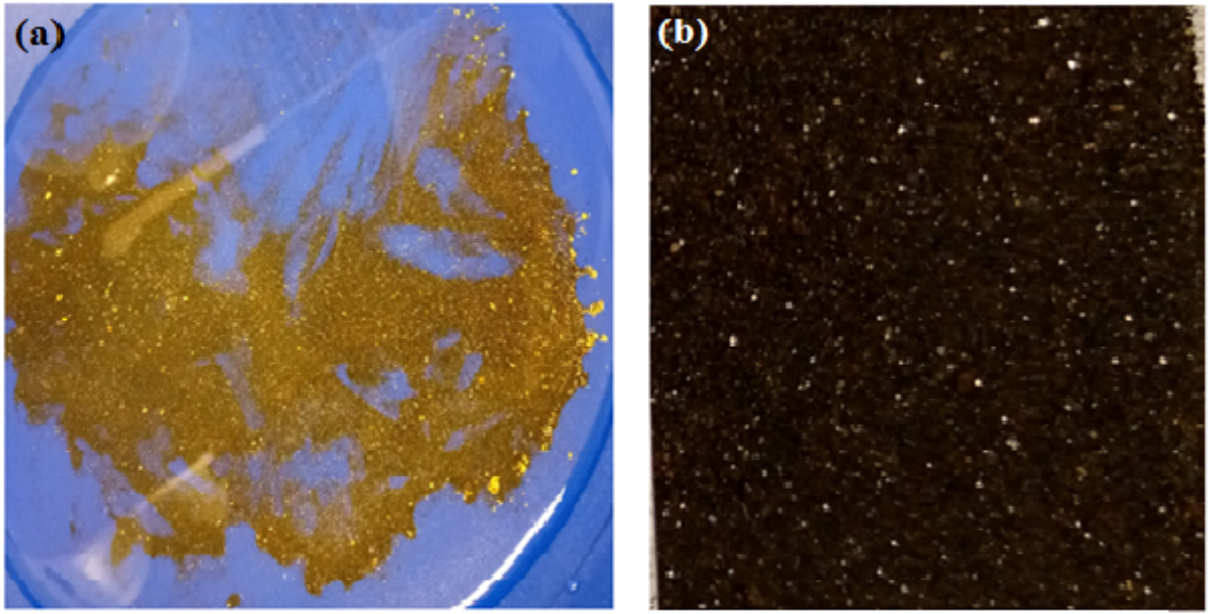

**Figure 3:** (a) Final residual sample containing Au, sand, and other magnetic materials to undergo magnetic separation (b) Impure $Fe_2O_3$ / $Fe_3O_4$ minerals. Photos by G. Nzulu.

### 2.4    X-ray diffraction measurement

The samples (both solid and powder forms) were irradiated using a PANAnalyical X'pert [25] powder diffractometer with a theta-2theta configuration. The operating conditions and equipment settings were Cu-K$\alpha$ radiation wavelength of 1.5406 Å ($\approx$ 8.04 keV); Cu long-fine-focus tube set to 45 kV and 40 mA; scan step size of 0.033; counting time of 10.16 s per step and scan range between 30 and 100 degrees in 2 theta scans. The size of the solid bulk Au nugget was 2x1.6x0.5 cm. The powder samples of impure Au, $Fe_2O_3$ and $Fe_3O_4$ had varying grain sizes (0.05-0.2 cm) and were put on a sample holder mounted on the diffractometer's sample mounting stage such that the crystal face was properly oriented and closely aligned with the diffractometer circle of the goniometer. The XRD data were quantitatively analyzed by Rietveld refinement using the MAUD software [26, 27].

### 3.    Results and Discussion

Figure 4 shows an x-ray diffractogram of the Au nugget sample in Fig. 2b with the result of Rietveld refinement assuming pure Au together with the residual of the fit [28]. The six pronounced peaks in the diffractogram are indexed as a cubic fcc Au structure (Fm-3m space group) with lattice parameter of $a$ = 4.079 Å. Table I lists the full assigned observed peak list as well as the resulting crystallographic parameters from the refinement. These are in agreement with literature assignments for Au [28, 29].

Figure 5 shows an x-ray diffraction of the impure powder Au sample shown in Fig. 3a. Table II lists the refined crystallographic parameters of Au and the pathfinder minerals identified from powder Au samples. These data are in agreement with reference data [27-32]. The diffractogram from the unrefined powder sample shows the presence of other minerals *i.e.*, pathfinder minerals for Au. These are dominated by $SiO_2$ (quartz) with some $Fe_3O_4$ (magnetite). The lattice parameter of the $SiO_2$ in the impure Au were found to be *a* = 4.91 Å and *c* = 5.43 Å (space group P3221), consistent with reference data [30]. This sample also contains $Fe_3O_4$ (cubic, space group Fd-3m) with a lattice parameter of 8.36 Å, consistent with literature data [27].

Figure 6 shows an x-ray diffractogram from the residual black sand after Au panning. The diffraction peaks of this sample were identified as the crystalline structure of $Fe_2O_3$ (hematite). Table III lists the diffraction peaks and crystallographic parameters determined from Rietveld refinement of $Fe_2O_3$. This is in accordance with literature and reference data [32, 33]. The crystal structure of $Fe_2O_3$ is rhombohedral with a space group R-3c and lattice constant of 5.0991 Å.

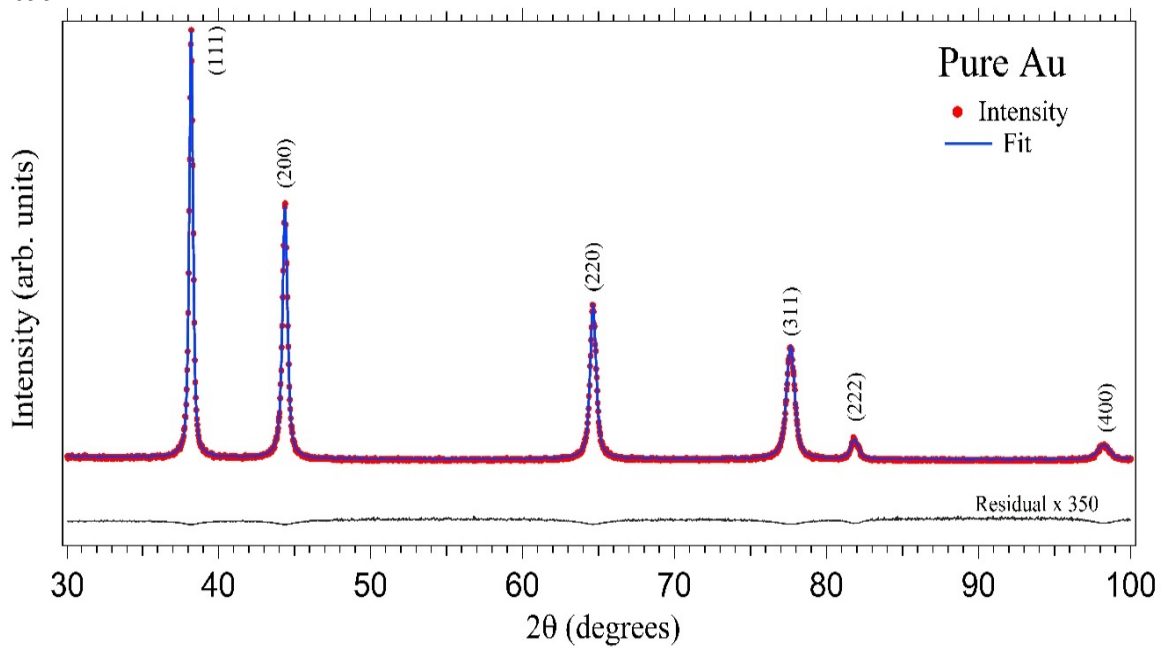

**Figure 4:** X-ray diffractogram of the bulk solid Au sample showing distinct peaks.

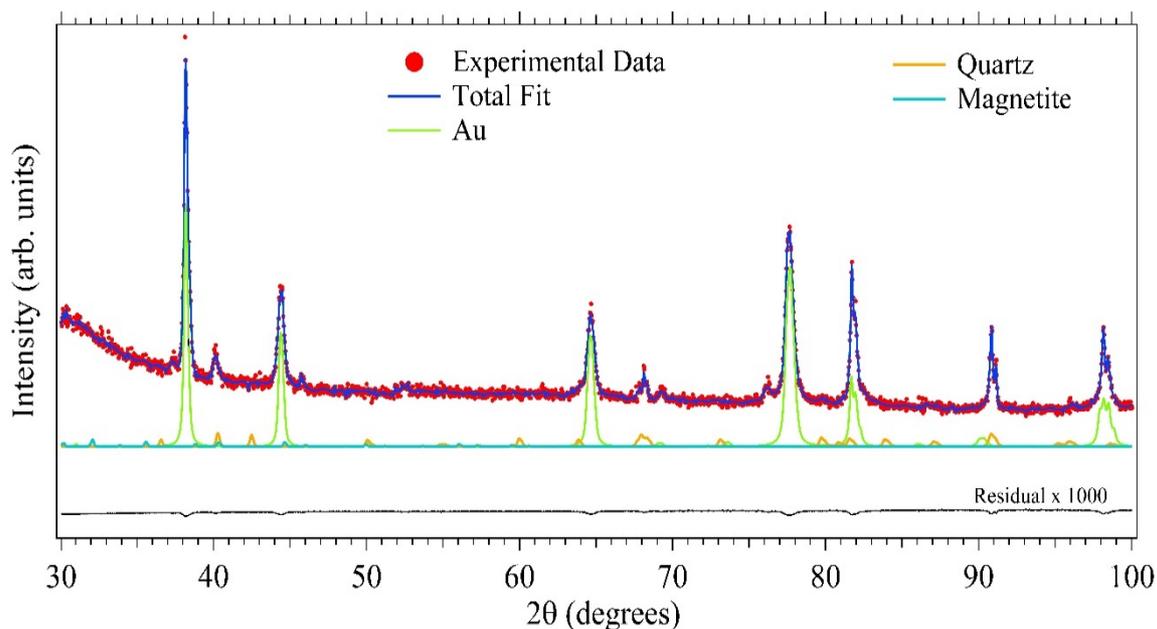

**Figure 5:** X-ray diffractogram of the impure powder Au with other pathfinding Au minerals.

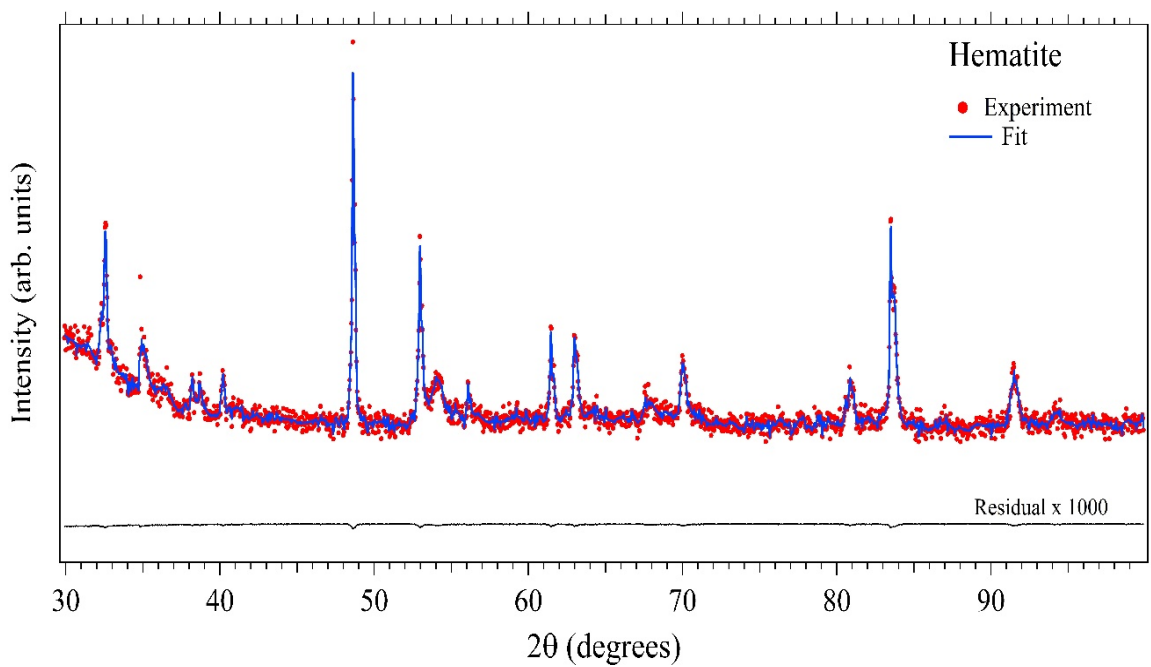

**Figure 6:** X-ray diffractogram of the $Fe_2O_3$ mineral.

**Table I:** Structural refinement parameters of solid bulk Au from XRD.
*********************************************************************

| | | |
|---|---|---|
| Symmetry: Cubic | Space group = Fm-3m | |
| Wavelength Cu Kα = 1.5406Å | COD ID: 9008463 | Ref. cell volume = 67.83 Å$^3$ |
| Wavelength Cu Kβ = 1.5444Å | | Refined cell volume = 67.82 Å$^3$ |

| OBSERVED | | | CALCULATED | | DIFFERENCE | |
|---|---|---|---|---|---|---|
| 2 theta | d | h k l | 2 theta | d | 2theta | d |
| 38.211 | 2.35344 | 1 1 1 | 38.185 | 2.35500 | 0.026 | 0.00156 |
| 44.394 | 2.03895 | 2 0 0 | 44.393 | 2.03900 | 0.001 | -0.00005 |
| 64.615 | 1.40276 | 2 2 0 | 64.578 | 1.44200 | 0.037 | -0.03924 |
| 77.616 | 1.22911 | 3 1 1 | 77.549 | 1.23000 | 0.067 | -0.00089 |
| 81.761 | 1.17696 | 2 2 2 | 81.724 | 1.17740 | 0.037 | -0.00044 |
| 98.238 | 1.01882 | 4 0 0 | 98.137 | 1.01960 | 0.101 | -0.00078 |

| Cell Parameters | Refinement | Reference | Error |
|---|---|---|---|
| a (Å) | 4.07803 ± 5.7603E-5 | 4.07825 | -0.00022 |
| b (Å) | 4.07803 ± 5.7603E-5 | 4.07825 | -0.00022 |
| c (Å) | 4.07803 ± 5.7603E-5 | 4.07825 | -0.00022 |
| Alpha (°) | 90.0000 | 90.0000 | |
| Beta (°) | 90.0000 | 90.0000 | |
| Gamma (°) | 90.0000 | 90.0000 | |

**Table II:** Structural refinement parameters of impure Au powder sample containing other pathfinder minerals.
*****************************************************************

**Table II (a)**    Impure Au parameters    [52.23%]    COD ID: 9008463
*****************************************************************

Initial symmetry: Cubic    Space group = Fm-3m    Ref. cell volume = 67.830 Å$^3$
Refined cell volume = 67.832 Å$^3$

| OBSERVED | | | CALCULATED | | DIFFERENCE | |
|---|---|---|---|---|---|---|
| 2 theta | d | h k l | 2 theta | d | 2theta | d |
| 38.178 | 2.35539 | 1 1 1 | 38.185 | 2.35500 | -0.007 | 0.00039 |
| 44.641 | 2.02824 | 2 0 0 | 44.393 | 2.03900 | 0.248 | -0.01076 |
| 64.615 | 1.44126 | 2 2 0 | 64.578 | 1.44200 | 0.037 | -0.00074 |
| 77.617 | 1.22910 | 3 1 1 | 77.549 | 1.23000 | 0.068 | -0.00009 |
| 81.728 | 1.17735 | 2 2 2 | 81.724 | 1.17740 | 0.004 | -0.00005 |
| 98.171 | 1.01934 | 4 0 0 | 98.137 | 1.01960 | 0.038 | -0.00026 |

| Cell Parameters | Refinement | Reference | Error |
|---|---|---|---|
| a (Å) | 4.07830 ± 4.7828E-4 | 4.07825 | 5.0 E-5 |
| b (Å) | 4.07830 ± 4.7828E-4 | 4.07825 | 5.0 E-5 |
| c (Å) | 4.07830 ± 4.7828E-4 | 4.07825 | 5.0 E-5 |
| Alpha (°) | 90.0000 | 90.0000 | |
| Beta (°) | 90.0000 | 90.0000 | |
| Gamma (°) | 90.0000 | 90.0000 | |

**Table II (b)**    SiO$_2$ structural parameters.    [33.89%]    COD ID: 1538064
*****************************************************************

Crystal system: Hexagonal    Space group = P3221    Ref. cell volume = 112.979 Å$^3$

Refined cell volume = 113.36 Å$^3$

| OBSERVED | | | CALCULATED | | DIFFERENCE | |
|---|---|---|---|---|---|---|
| 2 theta | d | h k l | 2 theta | d | 2theta | d |
| 40.284 | 2.23698 | 1 1 1 | 40.300 | 2.23613 | 0.026 | 0.00085 |
| 67.957 | 1.37829 | 2 1 2 | 67.744 | 1.38210 | 0.001 | -0.00381 |
| 90.818 | 1.08167 | 3 1 2 | 90.831 | 1.08155 | 0.067 | 0.00012 |

| Cell Parameters | Refinement | Reference | Error |
|---|---|---|---|
| a (Å) | 4.90970 ± 1.9E-3 | 4.91304 | -0.00334 |
| b (Å) | 4.90970 ± 1.9E-3 | 4.91304 | -0.00334 |
| c (Å) | 5.43023 ± 3.7E-3 | 5.40463 | 0.02557 |
| Alpha (°) | 90.00 ± 0.00 | 90.0000 | 0.0000 |
| Beta (°) | 90.00 ± 0.00 | 90.0000 | 0.0000 |
| Gamma (°) | 120.00 ± 0.00 | 90.0000 | 0.0000 |

**Table II (c)** $Fe_3O_4$ structural parameters. [13.88%]
*************************************************************

Crystal system: Cubic                     Space group = Fd-3m
COD ID: 9005813    Ref. cell volume = 583.816 Å$^3$    Refined cell volume = 584.277 Å$^3$

| OBSERVED | | | CALCULATED | | DIFFERENCE | |
|---|---|---|---|---|---|---|
| 2 theta | d | h k l | 2 theta | d | 2theta | d |
| 33.800 | 2.64979 | 1 0 4 | 33.153 | 2.70000 | 0.647 | -0.05020 |
| 35.571 | 2.52190 | 1 1 0 | 35.612 | 2.51900 | -0.041 | 0.00290 |
| 38.813 | 2.31830 | 0 0 6 | 39.277 | 2.29200 | -0.464 | 0.02630 |
| 40.317 | 2.23523 | 1 1 3 | 40.855 | 2.20700 | -0.538 | 0.02820 |
| 44.002 | 2.05620 | 2 0 2 | 43.519 | 2.07790 | 0.483 | -0.02170 |
| 49.843 | 1.82806 | 0 2 4 | 49.480 | 1.84060 | 0.363 | -0.01250 |
| 58.059 | 1.58740 | 0 1 8 | 57.590 | 1.59920 | 0.469 | -0.01180 |

| Cell Parameters | Refinement | Reference | Error |
|---|---|---|---|
| a (Å) | 8.36 ± 0.00 | 8.3578 | 0.0022 |
| b (Å) | 8.36 ± 0.00 | 8.3578 | 0.0022 |
| c (Å) | 8.36 ± 0.00 | 8.3958 | 0.0022 |
| Alpha (°) | 90.00 ± 0.00 | 90.0000 | 0.0000 |
| Beta (°) | 90.00 ± 0.00 | 90.0000 | 0.0000 |
| Gamma (°) | 90.00 ± 0.00 | 90.0000 | 0.0000 |

**Table III:** Structural refinement parameters of $Fe_2O_3$ powder sample.
*************************************************************

Crystal system: Rhombohedral              Space group = R-3c

COD ID: 900139  Ref. cell volume = 302.722 Å$^3$     Refined cell volume = 313.870 Å$^3$

| OBSERVED | | | CALCULATED | | DIFFERENCE | |
|---|---|---|---|---|---|---|
| 2 theta | d | h k l | 2 theta | d | 2theta | d |

| | | | | | | | |
|---|---|---|---|---|---|---|---|
| 32.609 | 2.74380 | 1 0 1 | | 33.158 | 2.70000 | -0.549 | 0.04380 |
| 34.915 | 2.56768 | 1 1 0 | | 35.612 | 2.51900 | -0.697 | -0.04868 |
| 38.658 | 2.32725 | 0 0 6 | | 39.277 | 2.29200 | -0.619 | 0.03525 |
| 40.196 | 2.24167 | 1 1 3 | | 40.855 | 2.20700 | -0.659 | 0.03467 |
| 48.618 | 1.87122 | 0 2 4 | | 49.480 | 1.84060 | -0.862 | 0.03062 |
| 53.966 | 1.69772 | 1 1 6 | | 54.091 | 1.69410 | -0.125 | 0.00362 |
| 62.990 | 1.47447 | 2 1 4 | | 62.451 | 1.48590 | 0.539 | -0.01143 |
| 69.975 | 1.34340 | 2 0 8 | | 69.601 | 1.34970 | 0.374 | -0.00630 |
| 80.837 | 1.18806 | 1 2 8 | | 80.711 | 1.18960 | 0.126 | -0.00154 |
| 83.475 | 1.15710 | 1 3 4 | | 84.916 | 1.14110 | -1.441 | 0.01600 |
| 91.533 | 1.07508 | 0 4 2 | | 91.345 | 1.07680 | 0.188 | -0.00172 |

| Cell Parameters | Refinement | Reference | Error |
|---|---|---|---|
| a (Å) | 5.0991 ± 1.4E-3 | 5.0380 | 0.0611 |
| b (Å) | 5.0991 ± 1.4E-3 | 5.0380 | 0.0611 |
| c (Å) | 14.0767 ± 5.2E-3 | 13.7720 | 0.3047 |
| Alpha (°) | 90.00 ± 0.00 | 90.0000 | 0.0000 |
| Beta (°) | 90.00 ± 0.00 | 90.0000 | 0.0000 |
| Gamma (°) | 120.00 ± 0.00 | 120.0000 | 0.0000 |

Comparing Figs 4 and 5, it can be seen that the latter sample contains Au together with pathfinder minerals in the form of magnetite and quartz. The most abundant mineral observed in the diffraction pattern of the impure powder Au sample is $SiO_2$ (quartz) having three distinct peaks at $2\theta = 40.284°$, $67.957°$ and $90.818°$ corresponding to {111}, {212} and {312} crystalline planes of the $SiO_2$ phase, respectively. The refined pattern of $SiO_2$ shown in Figure 5 is in agreement with the literature data in refs [36] [37] [38] [39], which also holds true for the moderate amount of magnetite present [27]. This shows that impure Au or final concentrate (non-pure Au) have a high quantity (percentage) of pathfinder minerals as impurities. Note that Au atoms easily substitutes with Ag atoms forming an alloy with the same fcc crystal structure and that it is impossible to distinguish pure Au from an Au-Ag alloy with XRD.

The diffractogram in Figure 6 contains major peaks at $2\theta = 32.609°$, $34.915°$, $38.658°$, $40.196°$, $48.618°$, $53.966°$, $62.990°$, $69.975°$, $80.837°$, $83.475°$ and $91.533°$ identified as {101}, {110}, {006}, {113}, {024}, {116}, {214}, {208}, {128}, {134} and {042} crystalline planes of $Fe_2O_3$ (hematite), respectively. These refined peaks are in good agreement with the rhombohedral structure of $Fe_2O_3$ [32] [33].

Generally, including possible microstrain in the Rietveld refinement has negligible effect for the convergence of the fit (residual), indicating that the samples are essentially strain-free. The results from the impure Au powder sample indicate that $SiO_2$ (quartz) is the dominant impurity mineral serving as the host rock containing all the pathfinder minerals at the mining site. It is known that $SiO_2$ is a so-called *gangue mineral* (*i.e.*, a commercially nonvaluable mineral that surrounds or is mixed with a valuable mineral) in hydrothermal ore veins [40], to preserve information about the physiochemical situations of the origin of the veins and to understand the formation of mineral deposits. These dominant $SiO_2$ species contain structural defects that favor mineral infusion due to underlying conditions and geological processes such as crystallization, metamorphism, alterations, changes in crystallization temperatures, and precipitation [41-43].

Au associated with $Fe_3O_4$ are mostly formed in skarns of granular magnetite usually found in contact with metamorphosed areas with magma intrusion into carbonate or silico-carbonate rocks that also consist of garnet and silicate minerals, among others. The residual black sand together with other dense minerals are considered to be ore that is left over during Au refinement and washing at riverbanks when recovering its Au content [44]. This shows that two of the three most common iron ore minerals; $Fe_3O_4$ and $Fe_2O_3$ are widely spread within the mining site and contribute to the Au host minerals alongside $SiO_2$. In a near surface environment (oxide area) $Fe_2O_3$ act as the gangue mineral and can be transformed to $Fe_3O_4$ depending on the environmental conditions such as high temperature, oxidation, and pH. The same color of $Fe_2O_3$ in comparison to black $Fe_3O_4$ makes it difficult to distinguish between the two in branded iron formations and in standing water [44, 46]. It is likely that during the formation of Fe-oxides in the alluvia regime at the Dunkwa-Kubi geological site, Au is internally captured within structures associated with $Fe_2O_3$ (hematite) that acts as crusts in saprolite and laterite environments. These minerals reveal information about the physiochemical conditions of the origin of structures (structural defects) useful for the understanding of mineral deposit formations.

## 4. Conclusions

This study has revealed that sediments and black sands containing Au are associated with pathfinding minerals in impure compositions. This is indicative that Au and pathfinding minerals are all deposited in nature during hydrothermal activation. The XRD analysis identified Au, $SiO_2$ (quartz), $Fe_3O_4$ (magnetite) and $Fe_2O_3$ (hematite). From the XRD patterns, the impure Au and $Fe_2O_3$ samples can be attributed to decomposition and transformation of these indicator minerals. Also, the surface (oxide zones) mineralization is altered by $Fe_2O_3$ as one of the indicator minerals apart from the garnet and the gangue mineral $SiO_2$ to host Au with other pathfinder minerals beneath the surface.
These results are of importance for the mining industry to underscore the usefulness of XRD in studying soil and sand sediments from mining sites by identifying pathfinder minerals of Au in potential geological sites.

## Acknowledgements

We acknowledge support from the Swedish Government Strategic Research Area in Materials Science on Functional Materials at Linköping University (Faculty Grant SFO-Mat-LiU No. 2009 00971). M.M. also acknowledges financial support from the Swedish Energy Research (Grant No. 43606-1) and the Carl Tryggers Foundation (CTS20:272, CTS16:303, CTS14:310). Asante Gold Corporation is acknowledged for funding G. K. N.'s industrial PhD studies at Linköping University, Sweden.

**Compliance with Ethical Standards:**
P. E. and M. M. declare no competing financial interest. G. K. N.'s industry PhD studies are funded by Asante Gold Corporation. Asante Gold Corporation or G.K.N. have no potential financial benefit from this study. The samples in this study are from an artisanal mining site open to the indigenous public.